\documentclass[allclo]{FBSart}
\usepackage{amsfonts}
\usepackage{amssymb}
\usepackage{graphicx}

\title{Universality and Leading Corrections in Few-Body Systems}
\author{L. Platter}
\institute{Department of Physics, Ohio State University, Columbus, OH 43120, USA}

\runningauthor{L.\,Platter}
\runningtitle{Style File for Few-Body Systems}
\newcommand{\beq}{\begin{equation}}
\newcommand{\eeq}{\end{equation}}
\newcommand{\bea}{\begin{eqnarray}}
\newcommand{\eea}{\end{eqnarray}}
\begin{document}

\maketitle
\begin{abstract}
A large two-body scattering length leads to universal behavior in few-body
systems. In particular, the three-body system displays interesting features
such as exact discrete scale invariance in the bound state
spectrum in the limit of infinite scattering length. Here, I will discuss how
an effective field theory (EFT) can be used to study these features and how the
finite range of the underlying interaction impacts the bound state spectrum
at first order in the EFT expansion.
\end{abstract}

\section{Introduction}
Few-body systems with a large two-body scattering length $a$ display
interesting universal features. In the two-body system a large positive two-body
scattering length will lead to a bound state with binding energy proportional
to $1/(M a^2)$ (where $M$ denotes the mass). Vitaly Efimov showed that the
situation in the three-body system is more complicated. For example, for
infinite scattering length the binding energies of
different states labeled with $n$ and $n_*$ are related by
\begin{equation}
B_3^{(n)}=(e^{-2\pi/s_0})^{n-n_*}B_3^{(n_*)}~,
\end{equation}
where $s_0\approx 1.00624$. The geometric spectrum is a signature of discrete
scaling symmetry with scaling factor $e^{-2\pi/s_0}$. Efimov pointed out
furthermore that these results are also relevant for finite scattering length
$a$ as long as $a \gg l$, where $l$ denotes the range of the underlying
interaction \cite{Efi71,Efi79}.

Over the last years an effective field theory (EFT) has been developed which
is tailored to calculate the low-energy properties of few-body systems with a
large two-body scattering length \cite{Braaten:2004rn}. This short-range EFT
is the appropriate description of ultracold atoms close to a Feshbach
resonance and nucleons at very low energies. At leading order, the short-range EFT
provides a powerful framework to calculate observables in the zero-range limit
and reproduces therefore the results derived by Efimov for the three-body
sector exactly. It allows furthermore to calculate the effects of the finite
range of the underlying interaction systematically and to compute electroweak
reactions relevant to nuclear astrophysics.

\section{One-Parameter Correlations and Universality}
A particular feature of the short-range EFT is the appearance of a three-body
force at leading order. Once this three-body counterterm is adjusted such that
a known three-body datum is reproduced all remaining observables can be
predicted.  Three-body observables will therefore depend not only on the
scattering length $a$ but also on one additional three-body parameter.

\begin{figure}[t]
\centerline{
\includegraphics[width=6.5cm,height=5cm,angle=0,clip=true]{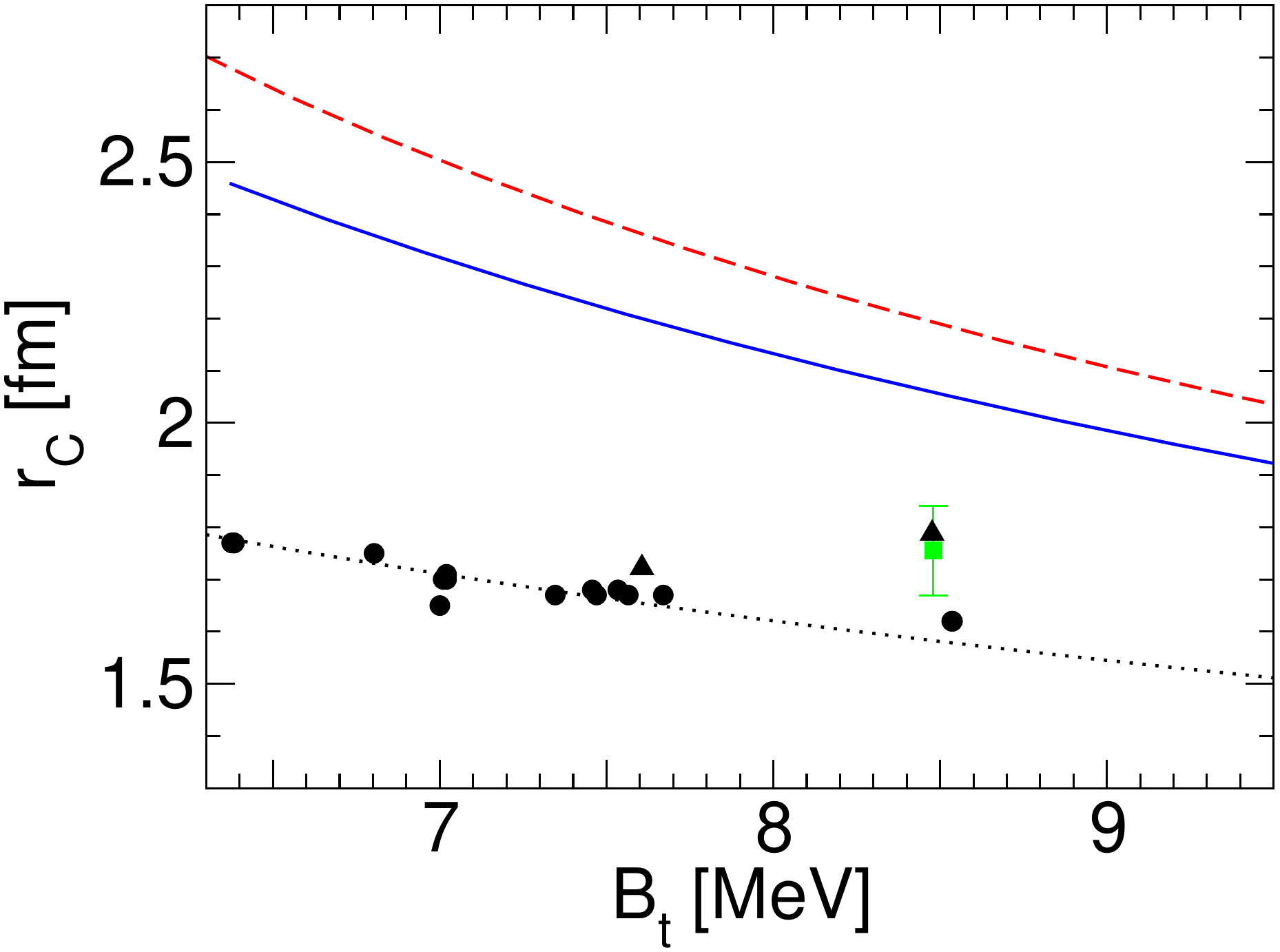}
\includegraphics[width=6.5cm,height=5cm,angle=0,clip=true]{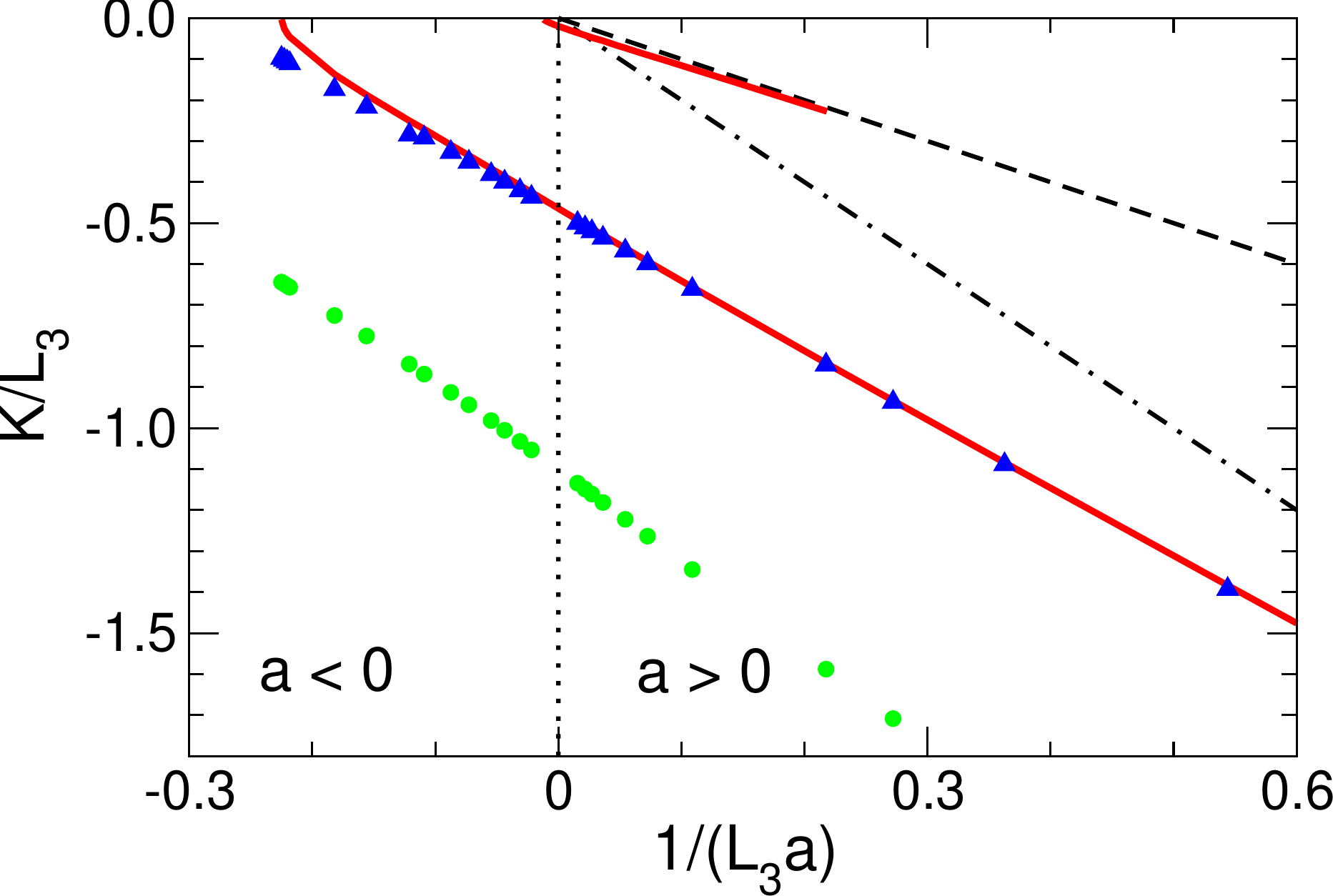}
}
\caption{\label{fig:opc}
Left panel: The correlation between the triton charge radius and binding energy. The
solid and dashed
lines denote the leading-order result using different two-body input parameters
The circles and triangles indicate Faddeev calculations using
different internucleon potentials. The square indicates the experimental
values.
Right panel:
The $a^{-1}-K$ plane for the four-body and three-body problems.
The circles and triangles indicate the four-body ground and excited state
energies, respectively , while the lower (upper) solid lines give the
thresholds for decay into a ground state (excited state) trimer and a particle. The dash-dotted
(dashed) lines give the thresholds for decay into two dimers (a dimer and two particles). The
vertical dotted line indicates infinite scattering length. All quantities are given in units of a
three-body parameter L3
}
\end{figure}
The necessity of this counterterm is more than just an artefact of the
field-theoretic formulation of the problem but is instead strongly tied to the
afore-mentioned discrete scale invariance in the three-body system.
Its appearance at leading order implies furthermore that two types of
one-parameter correlations can be generated within this framework.
Either the three-body counterterm is kept constant while the two-body scattering
length is varied or $a$ is kept constant while the
three-body counterterm is varied. 

One example for each of these types of one-parameter correlations is shown in
Fig.~\ref{fig:opc}. The correlation between the charge radius of the triton
and its binding energy is shown in the left panel of
Fig.~\ref{fig:opc}.  The two different solid lines correspond to two different
choices of fixing the two-body counterterms, the circles and triangles denote
Faddeev calculations with different potentials, the square denotes the
experimental value. The difference between the experimental result and the
short-range EFT result is due to the importance of finite range corrections.

The right panel shows an Efimov plot which includes results for the four-body
system obtained with the short-range EFT.  The circles and triangles denote
the binding momentum of two different four-body bound states which lie between
two successive three-body states.  The lower solid line denotes the shallower
of these two three-bound states.  The upper solid line denotes the next
three-body state in the Efimov spectrum.  The dashed and dot-dashed lines
denote the thresholds for decay into atom plus dimer and two dimers,
respectively.

\section{Finite Range Corrections}
A systematic calculation of higher order corrections is required for an
appropriate description of observables if the range of the underlying
interaction leads to a sizeable expansion parameter. This is the case in
nuclear physics where the ratio of effective range over scattering length is
roughly $\sim 1/3$.

Higher order corrections in the EFT expansion have been studied extensively over
the last years \cite{Hammer:2001gh,Bedaque:2002yg,Platter:2006ev}, however,
analytical information on the form of these range corrections is very limited.
In \cite{Platter:2008cx} we used the fact that the wave functions of the
Efimov trimers
are known in the unitary limit. This allowed us to calculate the shift in the
binding energies linear in the effective range in perturbation theory. It is
therefore necessary to calculate first the perturbing hyperspherical
potential \cite{Efimov91,Efimov93,Platter:2008cx}
\begin{equation}
 V_{\rm NLO}=-\frac{s_0^2\xi_0 r_s}{R^3}~,
\end{equation}
which is done by
implementing a next-to-leading order Bethe-Peierls condition into the
hyperangular equation. The shift in the binding energy of the $n$th bound
state can then be found by calculating the integral
\begin{equation}
  \frac{2 M}{\hbar^2} \Delta B^{(1)}_n=s_0^2 r_s \xi_0
  \left[\int_{\frac{1}{\Lambda}}^\infty dR {f_n}^2(R) \frac{1}{R^3} - \frac{2
      H_1 M}{\hbar^2 s_0^2 \,r_s \,\xi_0} \Lambda^2
    {f_n}^2\left(\frac{1}{\Lambda}\right)\right]~,
\end{equation}
where $f_n(R)$ is the
leading-order wave function of the $n$th three-body bound state. The second
term on the right hand side arises from a three-body force
\begin{equation}
V^{(1)}_{SR}(R)=H_1(\Lambda) \Lambda^2 \delta\left(R-\frac{1}{\Lambda}\right)
\end{equation}
which has been included to regularize the divergent first term. The
expression is renormalized by demanding that the shift in the binding energy
of the state with index $n_*$ is 0. It turns out that this condition leads
to the surprising result that the complete three-body spectrum
remains unperturbed, i.e. 
\begin{equation}
\Delta B^{(1)}_n=0~,
\end{equation}
for all $n$. This result which was also found numerically by Th\o gersen {\it
  et al.} \cite{thogerson2008} shows that the discrete scaling symmetry in the
three-body system constrains the form of higher order corrections strongly.
\section{Summary}
Effective field theories can be applied to any system in which a separation of
scales is present. They are not only perfectly suited to calculate observables
in a systematic low-energy expansion, but also provide a reliable error
estimate and a well-defined domain of applicability. An EFT appropriate for
short-range interactions has been applied to a large variety of physical
systems. I discussed how this short-range EFT can be used to study universal
relations in the three-body sector and how range corrections affect
the three-body bound state spectrum.

It is a surprising result that the Efimov spectrum in the unitary limit
remains unchanged at next-to-leading order. It will be interesting to see how
a finite range effects the universal relations between different three-body
observables such as the relation between the minima in the three-body
recombination rate and the binding energy of Efimov trimers in the unitary
limit. It might furthermore be possible to obtain analytic results at
next-to-next-to-leading order in the unitary limit.

The consistent inclusion of finite range corrections is required for future
calculations of electroweak reactions in few-body systems relevant to nuclear
astrophysics and will also be useful in applications of the short-range EFT to
Halo-nuclei \cite{Canham:2008jd} or $\alpha$-clusters \cite{Higa:2008dn}.
\begin{acknowledge}
This work was supported in part by the National Science 
Foundation under Grant No.~PHY--0653312,  
and the UNEDF SciDAC Collaboration under DOE Grant 
DE-FC02-07ER41457.
\end{acknowledge}

\end{document}